\documentclass[iop]{emulateapj}
\usepackage{color}
\usepackage{natbib}
\slugcomment{}

\shorttitle{Multiple stellar populations in NGC~2808}
\shortauthors{Carretta}

\begin{document}

\title{Five groups of red giants with distinct chemical composition in the
globular cluster NGC~2808\altaffilmark{1}}

\author{Eugenio Carretta\altaffilmark{2}}

\altaffiltext{1}{Based on data collected at the ESO telescopes under 
programme 072.D-0507 and during the FLAMES Science Verification programme.}
\altaffiltext{2}{INAF, Osservatorio Astronomico di Bologna, via Ranzani 1,
       40127,  Bologna,  Italy. eugenio.carretta@oabo.inaf.it}

\begin{abstract}
The chemical composition of multiple populations in the massive
globular cluster (GC) NGC~2808 is addressed with the homogeneous abundance
re-analysis of 140 red giant branch (RGB) stars. UVES spectra for 31 stars and
GIRAFFE spectra for the other giants were analysed with the same procedures used
for about 2500 giants in 23 GCs in our FLAMES survey, deriving abundances of Fe,
O, Na, Mg, Si, Ca, Ti, Sc, Cr, Mn, and Ni. Iron, elements from $\alpha-$capture,
and in the Fe-group do not show intrinsic scatter. On our UVES scale the
metallicity of NGC~2808 is  [Fe/H]$=-1.129\pm0.005\pm0.034$ ($\pm$statistical
$\pm$systematic error) with $\sigma=0.030$ (31 stars). Main features related to
proton-capture elements are retrieved, but the improved statistics and the
smaller associated internal errors allow to uncover five distinct groups of
stars along the Na-O anticorrelation. We observe large depletions in Mg,
anticorrelated with enhancements of Na and also Si, suggestive of unusually high
temperatures for proton-captures. About 14\% of our sample is formed by giants 
with solar or subsolar [Mg/Fe] ratios. Using the [Na/Mg] ratios we confirm  the
presence of five populations with different chemical composition, that we called
P1, P2, I1, I2, and E in order of decreasing Mg and increasing Na abundances.
Statistical tests show that the mean ratios in any pair of groups cannot be
extracted from the same parent distribution. The overlap with the five
populations recently detected from UV photometry is good but not perfect,
confirming that more distinct components probably exist in this complex GC.
\end{abstract}


\maketitle

\section{Introduction}

Galactic globular clusters (GCs) host multiple stellar populations of widely
(sometime wildly) different chemical composition and slightly different ages, as
shown since longtime by spectroscopy and more recently by photometric studies
(see the review by Gratton, Carretta, Bragaglia 2012 for a
summary of observations, references and topics on GCs). 

The first evidence from
abundance analyses in the last two decades of the past century revealed clear
C-N, Na-O anticorrelations, as well as the first correlations between Na and Al
abundances in a few GCs, reviewed by e.g. \citet{smi87} and \citet{kra94}.
Very
different abundances of these light elements in coeval stars at the same
evolutionary stage in otherwise monometallic stellar aggregates did negate the
very same definition of simple stellar population. However this concept was not
abandoned until two milestones were reached. First, theory provided the
understanding  that the observed abundance patterns could be generated by the
simultaneous action of the NaNa and MgAl cycles in the same stratification where
the ON part of the CNO cycle is operating in H-buning at high temperature
\citep[e.g.][]{den89,lan93}.
Second, the first
observation of the Na-O and Mg-Al anticorrelations among unevolved GC stars
\citep{gra01} unambigously demonstrated that the Na,Al-enhanced,
O,Mg-depleted composition  must be inherited by the most massive stars of a
(several) previous stellar generation(s).

Since then, the (anti)correlations among proton-capture elements have been
understood to represent the true DNA of GCs. These relations distinguish cluster
stars from the other {\it phylum} of Pop.II field stars where only the basic
evolutionary  effects - first dredge-up and a second mixing episode after the
red giant branch (RGB) bump in the luminosity function - on the lightest C and N
elements  are observed \citep{gra00}. The best studied chemical
signature is the Na-O anticorrelation, discovered by the Lick-Texas group 
\citep[see][]{kra94,sne00} and so widespread that can be considered as the
fingerprint of a  {\it bona fide} globular cluster \citep{car10a}. The
ubiquity of this important tracer shows that almost all GCs are composed of
multiple populations. 

The more recent photometric approach to multiple populations in GCs exploits the
presence in the filter passbands of absorption features of proton-capture
elements. In most cases only the lightest elements C,N are involved, with 
impact on those blue/near-UV filters including NH, CN, and CH features 
\citep[e.g.][and references therein]{gru98,mar08,han09,lee09,mon13,mil12a}.
Using
pseudo-color indexes, where stars of different generations are more or less
separated owing the different chemical composition, it is at present routinely
possible to uncover broad or split photometric sequences in most GCs. The recent
addition of the F275W passband of the WFC3 camera \citep{pio15}, which
samples an OH molecular band, is useful to connect these photometric sequences
with heavier proton-capture elements like O (and Na, through the NeNa and ON
cycles). Although no photometric filter seems to be presently able to detect
differences in the heaviest elements involved in proton-capture reactions (Mg,
Al, Si), space based UV photometry confirms that most GCs are formed by multiple
populations.

However, some GCs are more multiple than others. The most recent case is 
NGC~2808 where \citet{mil15} separated five distinct populations using
a combination of F275W, F336W, and F438W filters from $HST$. NGC~2808 is a
pivotal object to investigate in detail the complex scenario of the origin of
multiple populations in GCs, exploiting the wealth of photometric and
spectroscopic data gathered for this cluster. Apart from $\omega$ Cen and M~54,
whose metallicity spread tells a tale of a different origin (probably in/as a
core of a former dwarf nucleated galaxy), NGC~2808 is the fifth more massive
object in the catalogue by \citet{har96} of Galactic GCs ($M_V=-9.39$). The
present day total mass is found to be the driving parameter in establishing the
extent of chemical differences among multiple populations \citep{car10a}.
Moreover, the cluster mass is one of the more relevant factors
(together with the metallicity) in determining whether proton-capture reactions
are able to proceed to temperatures so high to significantly affect the
primordial level of heavier elements like Mg, from type II supernovae
\citep{car09a}. 

Three distinct main sequences (MS) were known since long time to exist in this
GC \citep{pio07,mil12b}, one of the best examples of the
impact on the colour-magnitude diagrams (CMDs) of helium, the main outcome of
the H-burning at high temperature. The distribution of stars along its
horizontal branch (HB) is multimodal \citep{bed00} and reach quite hot
effective temperature, again a good signature of enhanced He abundance in a
fraction of stars.

Our group is actively working since many years on the chemical composition of
stars in NGC~2808, gathering an unprecedented wealth of data. The first evidence
of multiple populations in this cluster came from the analysis of Na abundances
in 81 giants from moderately high resolution FLAMES spectra \citep{car03}. The 
spread in [Na/Fe]\footnote{We adopt the usual spectroscopic notation,
$i.e.$  [X]= log(X)$_{\rm star} -$ log(X)$_\odot$ for any abundance quantity X,
and  log $\epsilon$(X) = log (N$_{\rm X}$/N$_{\rm H}$) + 12.0 for absolute
number density abundances.} was found to span a range of about 1 dex similar to
what found in other GCs by the Lick-Texas group. Soon after, \citet{car04}
confirmed the multiple populations in NGC~2808 by finding an extended
Na-O anticorrelation on the RGB from high resolution UVES spectra of 20 stars,
with [O/Fe] ratios as low as -1 dex. This is not surprising, since there is a
tight correlation between the extent of the Na-O anticorrelation and the maximum
temperature along the HB \citep{car07a}, both features 
driven again by the same factor, He. NGC~2808 was one of the GCs used by 
\citet{carr06} to define the interquartile range of the [O/Na] 
ratio, IQR[O/Na], as an
efficient way to quantify the extension of the Na-O anticorrelation and the
relevance at large of changes in chemical composition due to the process of
formation of a GC and its multiple population.
Abundance analysis of horizontal branch (HB) stars was presented by 
\citet{gra11} and \citet{mar14}.

The analysis of GIRAFFE spectra for 120 giants in NGC~2808 \citep{car06}
started our FLAMES survey devoted to study the link between Na-O
anticorrelation and HB in about 25 GCs \citep{car06,car07b,car07c,car07d,
car09a,car09b,car10b,car11,car13a,car14,car15,gra06,gra07}.
However, being the first target considered in the survey, NGC~2808 was
analysed with slightly different procedures from those used for the other GCs,
in particular the scale of effective temperatures, efficiently tuned  for all
other GCs to minimize the impact of uncertainties in the atmospheric parameters
on star-to-star errors in abundances. This unconsistency calls for a re-analysis
of all the available material to derive the chemical composition of stars in
NGC~2808 in homogeneous way with respect to the other GCs in the survey.

First results of this re-analysis concerning the Al abundances (only available
for 31 RGB stars with UVES spectra) were already presented in \citet{carr14}
along with Mg abundances. The stars were found clustered into three discrete
groups with different chemical composition along the Mg-Al
anticorrelation. The fractions of stars in the three components were found in
excellent agreement with the number ratios computed by \citet{mil12b} on
the three MSs.

In the present paper we provide the full derivation of the atmospheric
parameters for all stars with available spectra from previous analyses. For 31
giants with UVES datat we  derived abundances of Fe, O, Na, Si, Ca, Ti, Sc, Cr,
Mn, Ni. Homogeneous abundances of Fe, O, Na are derived again for the 123 giants
with GIRAFFE spectra and new abundances of Mg, Si, Ca, Ti, Sc, Cr, and Ni were
obtained for the first time for this large set of stars.  Homogeneous abundance
ratios in our total sample extend the analysis of discrete components along the
RGB of NGC~2808.

The paper is organized as follows: in \S2 we describe the analysis method and 
the derivation of the atmospheric parameters with associated errors,  the
results for abundances, the Na-O and Na-Mg anticorrelations are  presented  in
\S3,  whereas the chemical tagging of multiple populations in NGC~2808 is 
discussed in \S4. Finally, in \S5 we summarise our findings.

\section{Analysis\label{ana}}

\subsection{Sample}

The observational material of the present re-analysis consists in UVES spectra
for 31 red giants and GIRAFFE spectra for 123 giants, all acquired with the
FLAMES instrument \citep{pas02} at the ESO VLT-UT2 telescope. We
used proprietary UVES data (ESO Programme 072.D-0507) for 12 stars analysed 
in \citet{car09a} and UVES spectra from the FLAMES Science
Verification for 19 stars whose analysis was done in \citet{car04} and
\citet{carr06}. These spectra have a resolution of R$\sim 47,000$ and a
spectral range between 4800 and 6800~\AA, with a small gap at about 5900~\AA.
Coordinates and magnitudes for all 31 stars of the UVES sample can be found in
\citet{carr14}.

The GIRAFFE spectra were taken with the high resolution grating and two setups,
HR11 ($R=24100$, wavelength range 5597-5840~\AA) and HR13 ($R=21900$, 
wavelength range 6120-6405~\AA). In this GIRAFFE sample, 32 stars were observed
with HR11 only, 28 with HR13 only and 63 have observations with both setups.
Coordinates and magnitudes for these giants are in \citet{car06}.
Details of the observations and data reduction are reported in the original
papers.

\begin{figure}
\centering
\includegraphics[scale=0.40]{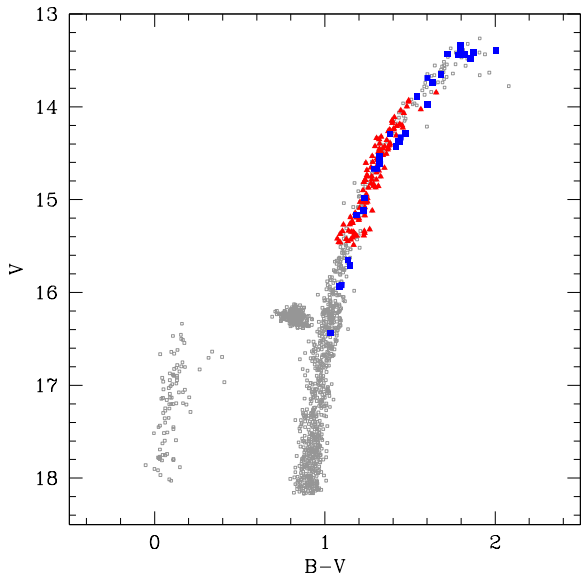}
\caption{$V,B-V$ CMD of NGC~2808 \citep[small squares]{bed00}. Stars
in the present sample observed with UVES are indicated with large blue squares,
giants observed with GIRAFFE are represented by large red triangles.}
\label{f:cmd2808}
\end{figure}

Fourteen out of 154 RGB stars were observed with both UVES and GIRAFFE, hence
our final sample consists of 140 individual stars, all members of the RGB of
NGC~2808 \citep[see the original papers by][]{car04,car06}, spanning the
magnitude range $V=13.34 \div 16.44$. In Fig.~\ref{f:cmd2808} the  location of
these RGB stars on the $V,B-V$ CMD from \citet{bed00} is  shown.
Only one star is fainter than the RGB-bump level $V=16.235$ \citep{nat11}.
The optical photometry was integrated with $K$ band
magnitudes from the Point Source Catalogue of 2MASS \citep{skr06} to
derive atmospheric parameters as described in the next Section.

\subsection{Atmospheric parameters and metallicity}

\begin{figure}
\centering
\includegraphics[scale=0.40]{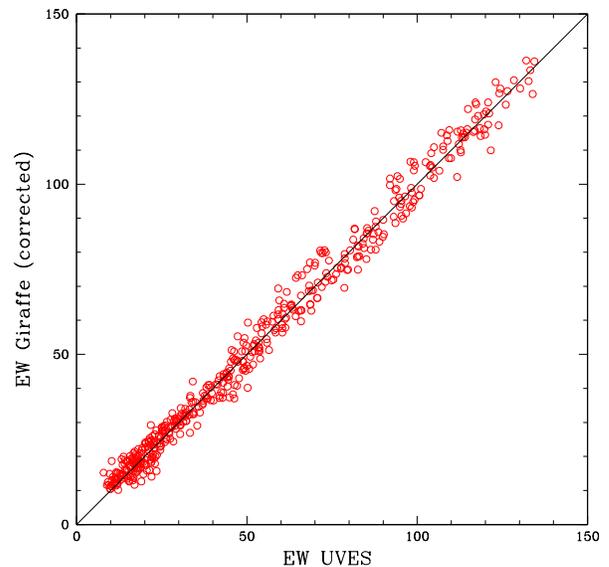}
\caption{ $EW$s measured on GIRAFFE spectra, after the correction to the UVES 
systems, as a function of the $EW$s measured on UVES spectra for the 14 giants
observed with both spectrographs.}
\label{f:ewcorr}
\end{figure}

All the abundances derived in the present re-analysis rest on equivalent widths
($EW$s) measured with the procedure described in \citet{bra01} and
using the latest version of the package ROSA \citep{gra88}. A first difference
with the original analysis in \citet{car06} is that this time the $EW$s
measured on the GIRAFFE spectra were corrected to the system defined by $EW$s from
the higher resolution UVES spectra. We used a linear regression between the two
sets of measurements derived using 478 lines in common
for the 14 stars observed with both instruments. The corrected $EW$s from
GIRAFFE are plotted as a function of the UVES $EW$s in Fig~\ref{f:ewcorr}.

This is the first step to make the present analysis perfectly consistent
with the abundance analysis of more than 2500 RGB stars in the other GCs
observed in our FLAMES survey.

The second step concerns the effective temperature (T$_{\rm eff}$) scale. In
\citet{car04,car06} and \citet{carr06} the giants of our sample were
analysed using T$_{\rm eff}$ values derived from $V-K$ colours and the
calibration by \citet{alo99,alo01}. The new approach followed for the
other 23 GCs and adopted here also for NGC~2808 is to use the above temperatures
as first pass values: the final values adopted for all our target stars
were obtained using an average relation between T$_{\rm eff}(V-K)$ and the $K$
apparent magnitudes of the RGB stars. This adopted procedure has three
advantages: it decreases the star-to-star scatter in abundances due to
uncertainties in temperatures (magnitudes are more precise than colours), it
minimises the impact of the differential reddening in NGC~2808 on derived
T$_{\rm eff}$, and finally it is fully consistent with the procedure adopted for
all the other GCs.

Surface gravities $\log g$ were obtained from bolometric corrections (from
Alonso et al.), the newly derived effective temperatures, reddening and distance
modulus from \citet{har96}, and assuming  masses of 0.85 M$_\odot$ and  
$M_{\rm bol,\odot} = 4.75$ as bolometric magnitude for the Sun.

Finally, values of the microturbulent velocity $v_t$ were obtained by 
eliminating in each star trends of the abundances from Fe~{\sc i} lines with 
the expected line strength (see Magain 1984). 
Final abundances were derived from $EW$ analysis, by choosing the model in the 
\citet{kur93} grid of solar-scaled LTE model atmospheres with the overshooting 
option switched on, with the proper atmospheric parameters, whose abundance
matched the one derived from Fe~{\sc i} lines. Adopted line list, atomic 
parameters and solar reference abundances are described in \citet{gra03}.

The new, final atmospheric parameters are listed with the derived Fe 
abundances of the 140 individual stars in Tab.~\ref{t:atmpart28}.

NGC~2808 is confirmed to be monometallic, as far the content of iron is
concerned. On the homogeneous metallicity scale based on high resolution UVES
spectra \citep{car09c} the metal abundance in NGC~2808 is on average
[Fe/H]~{\sc i}$=-1.129\pm0.005\pm0.034$ dex ($\sigma=0.030)$ from the 31 stars
observed with UVES (where the first and second error bars refer to the
statistical and systematic errors, respectively, see next Section). From the
reanalysis of the 123 stars with GIRAFFE spectra we derive on average 
[Fe/H]~{\sc i}$=-1.128\pm0.002\pm0.046$ dex ($\sigma=0.026)$. We then are
entitled to merge the two subsample, adopting the values from UVES for all
abundances in the 14 stars observed with both instruments.
The improvement due to the new homogeneous procedure is immediately evident in
the decrease of the $rms$ scatters associated to the mean Fe values, the 
previous values being $\sigma=0.075$ and $\sigma=0.065$ for UVES and GIRAFFE,
respectively.

Abundances of iron derived from singly ionized transitions are in very good
agreement with the above values. We derived 
[Fe/H]~{\sc ii}$=-1.128$ dex $(\sigma=0.026$ dex, 31 stars) and 
[Fe/H]~{\sc ii}$=-1.143$ dex $(\sigma=0.033$ dex, 91 stars) from the UVES and
GIRAFFE samples, respectively.

\begin{figure}
\centering 
\includegraphics[bb=78 171 460 691, clip,scale=0.52]{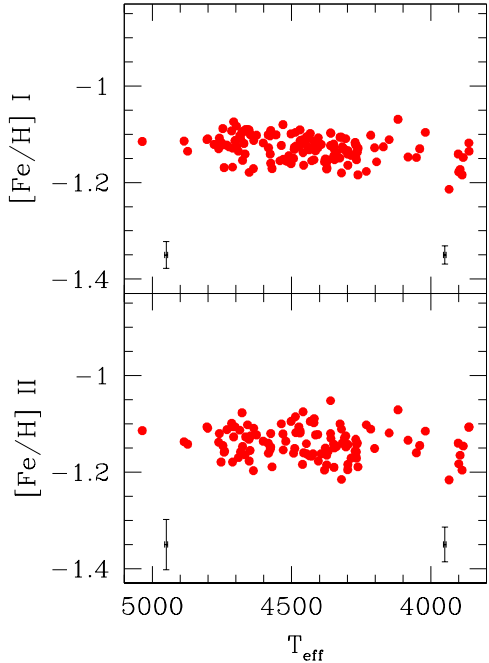}
\caption{Abundance ratios [Fe/H] {\sc i} (upper panel) and [Fe/H] {\sc ii}
(lower panel) as a function of T$_{\rm eff}$ for all 140 RGB stars in the
present re-analysis. Error bars on the right and on the left side are
star-to-star errors for targets observed with UVES and GIRAFFE, respectively.}
\label{f:feteff28}
\end{figure}

Derived abundances of [Fe/H]~{\sc i} and [Fe/H]~{\sc ii} are plotted as a
function of the effective temperatures in Fig.~\ref{f:feteff28}, together with
internal, star-to-star error bars as estimated in the next subsection.

\subsection{Abundances and error analysis}

Abundances of Fe, O, Na, Mg, Al, Si, Ca, Ti, Cr, Mn, and Ni for 19 giants with
UVES spectra from the FLAMES Science Verification were originally presented in
\citet{car04} and \citet{carr06}. O, Na, Mg, Al, and Si for the 12
RGB  stars observed with UVES in our FLAMES survey were analysed originally with
other 202 stars in 17 GCs in \citet{car09a}. These abundances are
derived again using the new scale of atmospheric parameters discussed above for
all 31 stars of the UVES sample and are presented here for the first time,
except for abundances of Al and Mg \citep{carr14}. For these stars, abundances
of Ti and Cr were obtained using both neutral and singly ionized species, with
results in excellent agreement.

For the 123 stars in the GIRAFFE sample, Fe, O, Na were revisited with respect
to the values in \citet{car06} obtained with the previous temperature
scale and without correcting the $EW$s to the UVEs system. Abundances of Mg, Si,
Ca, Ti, Cr, Ni are obtained for the first time in the present analysis. Due to
the limited spectral range of GIRAFFE spectra, only transitions of neutral Ti
and Cr were available for this set of giants.

Abundances of Na were corrected for effects of departures from the LTE
assumption using prescriptions by \citet{gra99}, whereas abundance
corrections for Sc and Mn as in \citet{gra03} were adopted to account
for the hyperfine structure.

In summary, we obtained homogeneous abundances of the elements O, Na, Mg, Si
involved in proton-capture reactions. Moreover, we derived abundances of
elements representative of the $\alpha-$capture process (Ca, Ti) and of the
Fe-group (Sc, Cr, Mn, Ni). Average abundance ratios and their $r.m.s.$ scatter 
are listed in Tab.~\ref{t:meanabu28} where the average value of Al by 
\citet{carr14} is also indicated. The abundances are listed separately 
for the UVES and
GIRAFFE sets, but whenever possible we adopted UVES abundances for individual
stars  in our merged sample.

\setcounter{table}{1}
\begin{table}
\centering
\caption{Mean abundances from UVES and GIRAFFE }
\begin{tabular}{lcc}
\hline
 lement              & UVES	     & GIRAFFE       \\
                     &n~~   avg~~  $rms$ &n~~	avg~~  $rms$ \\        
\hline
$[$O/Fe$]${\sc i}    &31   +0.06 0.34 &91   +0.00 0.38  \\
$[$Na/Fe$]${\sc i}   &31   +0.20 0.20 &123  +0.24 0.23  \\
$[$Mg/Fe$]${\sc i}   &31   +0.26 0.16 &122  +0.27 0.16  \\
$[$Al/Fe$]${\sc i}   &31   +0.46 0.47 & 	        \\
$[$Si/Fe$]${\sc i}   &31   +0.29 0.07 &123  +0.30 0.05  \\
$[$Ca/Fe$]${\sc i}   &31   +0.33 0.02 &123  +0.33 0.03  \\
$[$Sc/Fe$]${\sc ii}  &31 $-$0.01 0.04 &123$-$0.00 0.05  \\
$[$Ti/Fe$]${\sc i}   &31   +0.21 0.02 &123  +0.22 0.04  \\
$[$Ti/Fe$]${\sc ii}  &29   +0.18 0.02 & 	        \\
$[$Cr/Fe$]${\sc i}   &31 $-$0.04 0.02 &123$-$0.03 0.03  \\
$[$Cr/Fe$]${\sc ii}  &31 $-$0.03 0.03 & 	        \\
$[$Mn/Fe$]${\sc i}   &31 $-$0.37 0.04 & 	        \\
$[$Fe/H$]${\sc i}    &31 $-$1.13 0.03 &123$-$1.13 0.03  \\
$[$Fe/H$]${\sc ii}   &31 $-$1.14 0.03 &91 $-$1.14 0.03  \\
$[$Ni/Fe$]${\sc i}   &31 $-$0.06 0.02 &123$-$0.08 0.02  \\
\hline
\end{tabular}
\begin{list}{}{}
\item[] Note: individual abundances of Al are reported in \citet{carr14}.
\end{list}
\label{t:meanabu28}
\end{table}

The sensitivity of the derived abundances to variations in the adopted
atmospheric parameters for each element was obtained by re-iterating the
analysis and varying each time only one parameter of the amount shown in 
Tab.~\ref{t:sensitivityt28}. For the final slopes of the relations between the 
variation in each parameter and the abundance we took the average over all
stars. This computation was done separately for UVES and GIRAFFE.

These sensitivities were used to estimate the impact of uncertainties in
atmospheric parameters on the derived abundances. A detailed description is
given in \citet{car07a,car09a} respectively for the analysis of GIRAFFE
and UVES spectra. We note here that with respect to the analysis of NGC~2808 in
\citet{car06,car09a} the internal error in temperature is now reduced by
almost an order of magnitude, from 44 K to 5 K, with the presently adopted 
procedure for T$_{\rm eff}$. This improvement is very important to minimise the
star-to-star errors in abundances (Tab.~\ref{t:sensitivityt28}), a crucial step
for a meaningful study of  possible segregation of stars into groups with
discrete chemical composition.

\begin{table*}
\centering
\caption[]{Sensitivities of abundance ratios to variations in the atmospheric
parameters and to errors in the equivalent widths, and errors in abundances for
stars of NGC~2808 observed with UVES and GIRAFFE}
\begin{tabular}{lrrrrrrrr}

\multicolumn{9}{c}{UVES}      \\
\hline
Element     & Average  & T$_{\rm eff}$ & $\log g$ & [A/H]   & $v_t$    & EWs     & Total   & Total      \\
            & n. lines &      (K)      &  (dex)   & (dex)   &kms$^{-1}$& (dex)   &Internal & Systematic \\
\hline        
Variation&             &  50           &   0.20   &  0.10   &  0.10    &         &         &            \\
Internal &             &   5           &   0.04   &  0.03   &  0.05    & 0.01    &         &            \\
Systematic&            &  46           &   0.06   &  0.03   &  0.01    &         &         &            \\
\hline
$[$Fe/H$]${\sc  i}& 71 &    +0.036     &   +0.015 &  +0.003 & $-$0.031 & 0.009  &0.019    &0.034	\\
$[$Fe/H$]${\sc ii}&  9 &  $-$0.055     &   +0.102 &  +0.031 & $-$0.015 & 0.027  &0.036    &0.059	\\
$[$O/Fe$]${\sc  i}&  2 &  $-$0.022     &   +0.067 &  +0.029 &   +0.028 & 0.057  &0.061    &0.066	\\
$[$Na/Fe$]${\sc i}&  4 &    +0.011     & $-$0.048 &$-$0.034 &   +0.013 & 0.040  &0.043    &0.040	\\
$[$Mg/Fe$]${\sc i}&  3 &  $-$0.009     & $-$0.013 &$-$0.006 &   +0.016 & 0.046  &0.047    &0.029	\\
$[$Al/Fe$]${\sc i}&  2 &    +0.007     & $-$0.023 &$-$0.011 &   +0.021 & 0.057  &0.058    &0.086	\\
$[$Si/Fe$]${\sc i}&  7 &  $-$0.050     &   +0.020 &  +0.008 &   +0.021 & 0.030  &0.033    &0.048	\\
$[$Ca/Fe$]${\sc i}& 15 &    +0.027     & $-$0.031 &$-$0.015 & $-$0.017 & 0.021  &0.024    &0.027	\\
$[$Sc/Fe$]${\sc ii}& 8 &    +0.048     & $-$0.023 &$-$0.004 & $-$0.016 & 0.028  &0.030    &0.045	\\
$[$Ti/Fe$]${\sc i}&  9 &    +0.051     & $-$0.019 &$-$0.016 & $-$0.005 & 0.027  &0.028    &0.047	\\
$[$Ti/Fe$]${\sc ii}& 9 &    +0.041     & $-$0.030 &$-$0.010 & $-$0.024 & 0.027  &0.030    &0.039	\\
$[$Cr/Fe$]${\sc i}& 17 &    +0.036     & $-$0.030 &$-$0.015 & $-$0.007 & 0.019  &0.021    &0.035	\\
$[$Cr/Fe$]${\sc ii}&10 &    +0.016     & $-$0.026 &$-$0.017 &   +0.004 & 0.025  &0.026    &0.018	\\
$[$Mn/Fe$]${\sc i}&  3 &    +0.016     & $-$0.012 &$-$0.007 & $-$0.015 & 0.046  &0.047    &0.016	\\
$[$Ni/Fe$]${\sc i}& 26 &  $-$0.011     &   +0.015 &  +0.006 &   +0.010 & 0.016  &0.017    &0.012	\\

\hline
\hline

\\
\multicolumn{9}{c}{GIRAFFE}      \\

\hline
Element     & Average  & T$_{\rm eff}$ & $\log g$ & [A/H]   & $v_t$    & EWs     & Total   & Total      \\
            & n. lines &      (K)      &  (dex)   & (dex)   &kms$^{-1}$& (dex)   &Internal & Systematic \\
\hline        
Variation&             &  50           &   0.20   &  0.10   &  0.10    &         &         &            \\
Internal &             &   5           &   0.04   &  0.03   &  0.08    & 0.01    &         &            \\
Systematic&            &  46           &   0.06   &  0.05   &  0.01    &         &         &            \\
\hline
$[$Fe/H$]${\sc  i}& 37 &    +0.050     &   +0.005 &$-$0.003 & $-$0.030 & 0.013  &0.028    &0.046	\\
$[$Fe/H$]${\sc ii}&  3 &  $-$0.039     &   +0.096 &  +0.027 & $-$0.013 & 0.047  &0.052    &0.046	\\
$[$O/Fe$]${\sc  i}&  2 &  $-$0.041     &   +0.079 &  +0.035 &   +0.035 & 0.057  &0.067    &0.060	\\
$[$Na/Fe$]${\sc i}&  3 &  $-$0.004     & $-$0.037 &$-$0.019 &   +0.016 & 0.047  &0.049    &0.024	\\
$[$Mg/Fe$]${\sc i}&  2 &  $-$0.014     & $-$0.009 &$-$0.004 &   +0.016 & 0.057  &0.058    &0.020	\\
$[$Si/Fe$]${\sc i}&  8 &  $-$0.051     &   +0.026 &  +0.012 &   +0.026 & 0.029  &0.036    &0.048	\\
$[$Ca/Fe$]${\sc i}&  5 &    +0.015     & $-$0.029 &$-$0.010 & $-$0.019 & 0.036  &0.040    &0.016	\\
$[$Sc/Fe$]${\sc ii}& 5 &  $-$0.055     &   +0.077 &  +0.031 &   +0.001 & 0.036  &0.041    &0.030	\\
$[$Ti/Fe$]${\sc i}&  4 &    +0.032     & $-$0.013 &$-$0.012 &   +0.006 & 0.041  &0.041    &0.056	\\
$[$Cr/Fe$]${\sc i}&  5 &    +0.014     & $-$0.014 &$-$0.007 &   +0.022 & 0.036  &0.040    &0.014	\\
$[$Ni/Fe$]${\sc i}&  8 &  $-$0.014     &   +0.017 &  +0.008 &   +0.021 & 0.029  &0.033    &0.014	\\

\hline
\end{tabular}
\label{t:sensitivityt28}
\end{table*}

\section{Results}

For individual stars, abundances of  proton-capture elements are given in
Tab.~\ref{t:protont28}. In  Tab.~\ref{t:alphat28} are listed abundances of
$\alpha=$ and Fe-group  elements for species measured both on UVES and GIRAFFE
spectra. Abundances of Ti~{\sc ii}, Cr~{\sc ii} and Mn (only available from UVES
spectra) are listed in Tab.~\ref{t:ticrmnt28}. The run of abundance ratios in
NGC~2808  does not show any trend as a function of effective temperature.

\begin{figure}
\centering 
\includegraphics[scale=0.40]{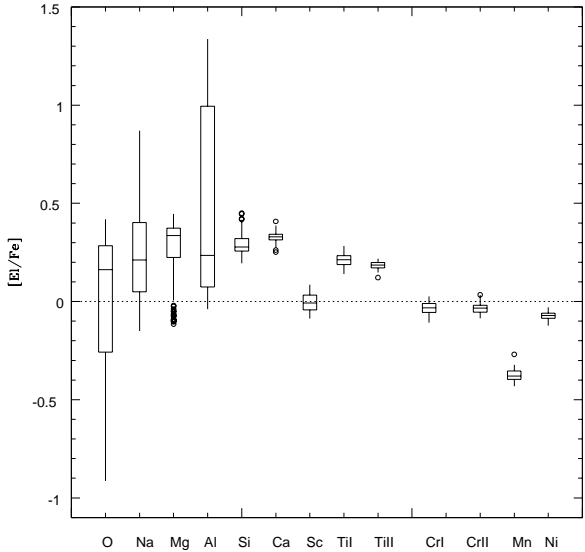}
\caption{Box-and-whiskers plot of abundance ratios in RGB stars of NGC~2808. The
horizontal line in each box is the median ratio of a given element. 
The bottom and the top lines of the box are the 25th and the 75th
percentiles, whose distance  (i.e. the box length) indicates the interquartile
range, encompassing the  middle 50\% of the data. The vertical ``whiskers"
protruding from the box indicate the full abundance range excluding the
outliers, defined as points lying more than 1.5 times the interquartile range
from the 25th or 75th percentiles and indicated with open circles.}
\label{f:figmed28}
\end{figure}

We provide in Fig.~\ref{f:figmed28} a graphical summary of the derived
abundances in NGC~2808, using the so called box-and-whishers plot \citep{tuk77}
which is particularly well suited to visualise possible skewed distributions and
to investigate the presence of outstanding outliers. 

The abundance pattern observed in NGC~2808 is typical of massive, normal
GCs. Light elements involved in proton-capture reactions present a huge spread,
exceeding many times the uncertainties associated to the analysis. Pure
$\alpha-$process elements like Ca and Ti (not touched by proton-capture
reactions, like instead Mg and Si) show a much more limited range, with
overabundances typical of normal nucleosynthesis from type II supernovae.
Similarly, species belonging to the Fe-peak follow rather well the run of
iron, apart from Mn which is underabundant as in normal halo stars of similar
metallicity.

However, already from this compact representation it is immediately clear that
Mg in NGC~2808 is outstanding even among proton-capture elements. The improved
statistics of the present analysis, 140 RGB stars compared to the maximum of 31
stars with Mg abundance analysed in \citet{carr14}, allows us to uncover an
unusually large number of outliers in the Mg distribution, all of them with
subsolar [Mg/Fe] ratios. Also the Si distribution seems to be particular, with a
number of outliers located at the high-abundance border of the distribution,
i.e. anti-correlated with the behaviour observed in Mg.

\subsection{The classical Na-O anticorrelation}

\begin{figure}
\centering 
\includegraphics[scale=0.40]{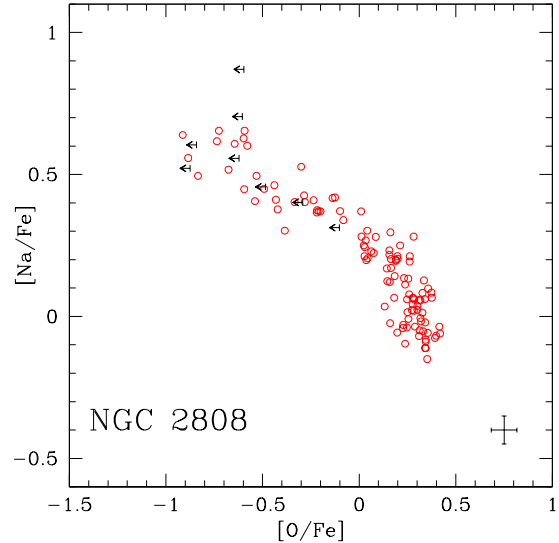}
\caption{Anticorrelation of O and Na abundances in RGB stars of NGC~2808
from the present reanalysis. Arrow indicate upper limits in [O/Fe]. Star-to-star
errors are those referred to the bulk of stars, observed with GIRAFFE. Internal
errors for the 31 stars with UVES spectra are slighly smaller.}
\label{f:m28antiu2}
\end{figure}

The updated version of the Na-O anticorrelation, the most outstanding typical
signature of the chemical composition in a GC, is shown in
Fig.~\ref{f:m28antiu2}. Abundances of Na are available for all 140
stars in our merged sample, because Na lines fall in both the HR11 and HR13
setups (and obviously in the UVES spectral range). However, abundances of O are
only available for 91 stars observed with the GIRAFFE setup HR13; when merged
with the UVES sample, our final sample includes 117 stars with both O and Na,
which is however the largest to date used to study this feature in a normal,
monometallic GC.

The present re-analysis confirms many previous results \citep{car06}:
the Na-O anticorrelation is very extended, showing stars with O abundances
reaching almost [O/Fe]$=-1$ dex. The interquartile range of the [O/Na] ratio,
IQR[O/Na], introduced by \citet{carr06} to provide a quantitative measure of
changes in chemical composition between stars of different generations, is
0.925. This value, although slightly smaller than the old value (0.999),
perfectly fits the relation with the cluster total present-day mass (as
represented by the proxy of the total absolute magnitude $M_V$). In the same
way, we confirm that NGC~2808 lies very well on the relation discovered by
\citet{car07e} linking the extension of the Na-O anticorrelation with
the HB morphology, as represented by the maximum temperature along the HB
\citep{rec06}.

The division of stars into the three components defined in \citet{car09b},
primordial P and two components of second generation stars with
intermediate I and extreme E composition, gives the fractions P$=46 \pm 6\%$,
I$=36 \pm 6\%$, and E$=18 \pm 4\%$, where the attached error are from Poisson
statistics. We confirm that NGC~2808 is one of the GCs where the highest 
fraction of first generation stars was retained: almost half of the present day
stellar population belongs to the P component, whereas the observed mean over
more than 20 GCs homogeneously analysed by our group is that only a third of 
stars show a pristine composition.   

\begin{figure}
\centering 
\includegraphics[scale=0.40]{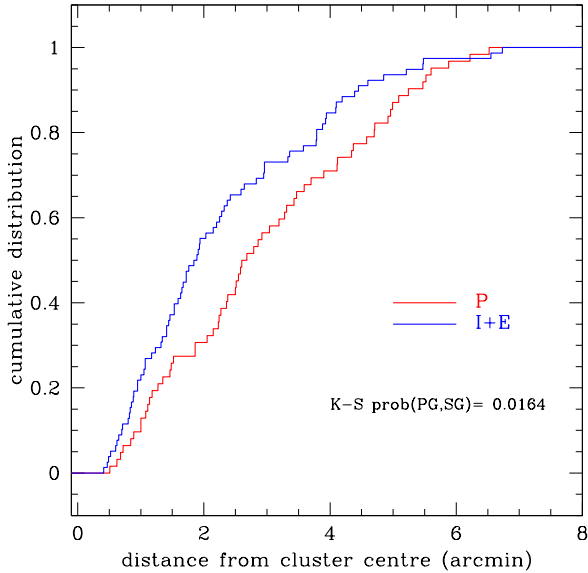}
\caption{Cumulative distribution of radial distances from the cluster centre for
stars of first (P component) and of second (I+E) generation in NGC~2808.
The probability of a Kolmogorov-Smirnov test is also listed.}
\label{f:distrFGSG28}
\end{figure}

Second generation stars (component I and E) show a tendency to be more 
concentrated than stars with primordial composition (Fig.~\ref{f:distrFGSG28}):
a two-tail Kolmogorov-Smirnov test allow us to safely reject the null hypothesis
that the two distributions are extracted from the same parent population.
However, due to the limitations imposed by the fiber positioning of FLAMES,
only a small fraction of stars in our sample is located within two hal-mass
radii from the cluster centre. Coupled to the different concentration of first
and second generation stars this suggests that more ample photometric dataset
are better suited for a more thorough study of this issue.

\begin{figure}
\centering 
\includegraphics[scale=0.40]{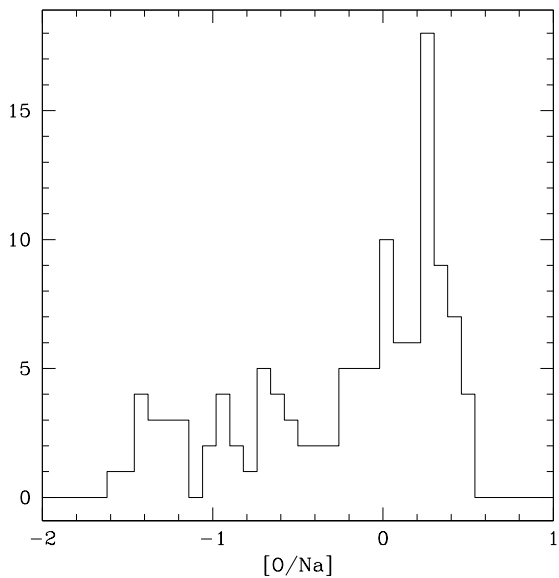}
\caption{Distribution function of the [O/Na] ratios of 117 giants in NGC~2808}
\label{f:histonao28}
\end{figure}

At first blush there seems to be not much difference with respect to previous 
results. However, the improved statistics of the merged sample and the decrease
of internal errors due to the atmospheric parameters uncover a more subtle 
level of complexity in this cluster. 
The distribution function of the [O/Na] ratio reveals that stars along
the Na-O anticorrelation seem to be splitted not simply in three, but into 
several structures, namely five
groups peaked at about [O/Na]$=+0.30, 0.0, -0.65, -0.9,$ and $-1.3$ dex (see
Fig.~\ref{f:histonao28}).

Let us assume, as a working hypothesis, that both the P and I components in
NGC~2808 are composed each of two sub-groups. The separation of the last three
peaks at low [O/Na] values in Fig.~\ref{f:histonao28} is quite  large, but even considering
conservatively the larger errors estimated for GIRAFFE also the distance  
between the first two peaks exceeds by almost four times the combined internal 
uncertainty in O and Na (see Tab.~\ref{t:sensitivityt28}). This means that the
ensemble of stars with primordial composition in NGC~2808 is not a monolitic
group, but rather composed by two sub-components, with small differences in the
average [O/Fe] ratio and larger differences in the Na  content.

The existence of these two groups is in excellent agreement with the independent
result obtained from the abundance analysis of HB stars in NGC~2808 by 
\citet{gra11}.
By observing several tens of stars they found some evidence of
Na-O anticorrelation already among the red HB (RHB) stars that should be the 
descendents of He-poor, O-rich stars. \citet{gra11} concluded that the
stellar populations in NGC~2808 must be more than three, since at least part of
the red HB stars does not belong to the primordial cluster population as
expected from the classical paradigm relating O-rich stars to normal helium
content.
Recently, \citet{mar14}, who did not derive O abundances for their sample
of HB stars, confirmed a bimodality in Na among RHB stars in NGC~2808.

We may tentatively conclude that {\it we found in our sample of RGB stars the
progenitors of the two groups that end up on the RHB, with normal He, 
similarly high O, and slightly different Na content}.

Two other groups seen in Fig.~\ref{f:histonao28} present a composition typical
of intermediate second generation stars, while the last is clearly identifiable
with the extreme component as defined in \citet{car09b}, characterised
by very low O abundances and high Na values.

The coincidence of five groups distinct by their different chemical composition
with  five populations found photometrically by \citet{mil15} is a result
beyond our expectations when this re-analysis was started simply  to bring
NGC~2808 onto the homogeneous system used for all other GCs in our FLAMES
survey.
Fortunately, we have now the possibility to investigate this finding in deeper 
details using abundances of other proton-capture elements as a sanity check,
showing that there is not a simple correspondence between results from
spectroscopy and photometry. The scenario in NGC~2808 looks more complex.

\subsection{The Na-Mg anticorrelation confirms five discrete groups in NGC~2808 }

Several reasons prompted us to use the relation between Na and Mg abundances  to
check the reality of the five chemically distinct components on the RGB of 
NGC~2808. These elements are the outcome of the two different cycles (NeNa and 
MgAl), hence they sample different reactions occurring at different 
temperatures, thus mass stratifications.  Their interplay is then a good
candidate to offer a panoramic view over all the mass range of possible
polluters that contributed gas enriched in proton-capture elements at the epoch
of cluster formation.

From a more practical point of view, all 140 giants in our final sample have Na
abundances and we cannot derive a Mg abundance only for one star, with a net
gain of about 20\% in statistics with respect to the O-Na plot.

\begin{figure}
\centering 
\includegraphics[scale=0.40]{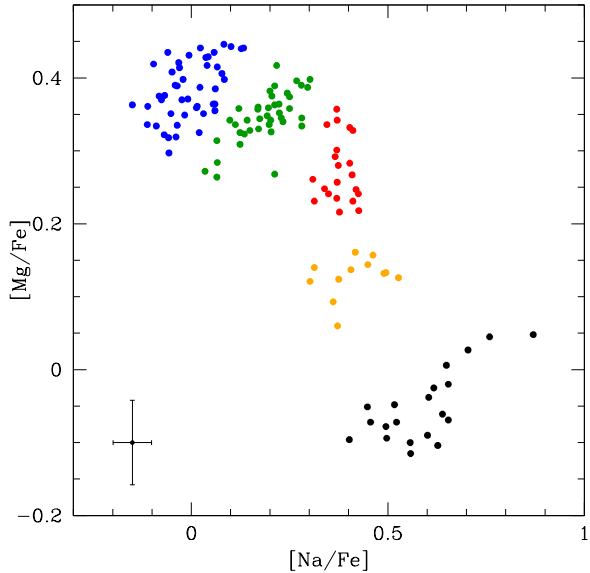}
\caption{Anticorrelation between abundances of Mg and Na in RGB stars of
NGC~2808. Different colours are used to indicate giants of different groups.
Internal error bars are those derived for GIRAFFE, the ones relative to UVES
being slightly smaller.}
\label{f:gruppi}
\end{figure}

Abundances of Na and Mg are plotted in Fig.~\ref{f:gruppi}. The ratios [Na/Fe]
and [Mg/Fe] in RGB stars of NGC~2808 are anticorrelated, with Mg decreasing by
about 0.55 dex while Na increases by about 1 dex. A Spearman correlation test
provides a probability that Na and Mg are not anticorrelated of less than 
$1.0 \times 10^{-6}$.
This feature is not surprising, apart from the large fraction of stars with
severe Mg depletions in this GC, being simply the consequence of Na production
from $^{22}$Ne with simultaneous Mg consumption from the MgAl cycle. However,
the key feature we observe in NGC~2808 is that the stars appear not to be
uniformly distributed along the anticorrelation, but instead it looks like they are
clustered in several distinct groups. In analogy with the division based on O,Na
abundances we call these five groups P1, P2 (with primordial or almost
primordial abundance ratios), I1, I2 (with intermediate composition), and E
(with severely extreme changes from the original composition).

\begin{figure*}
\centering 
\includegraphics[bb=19 146 579 461, clip, scale=0.52]{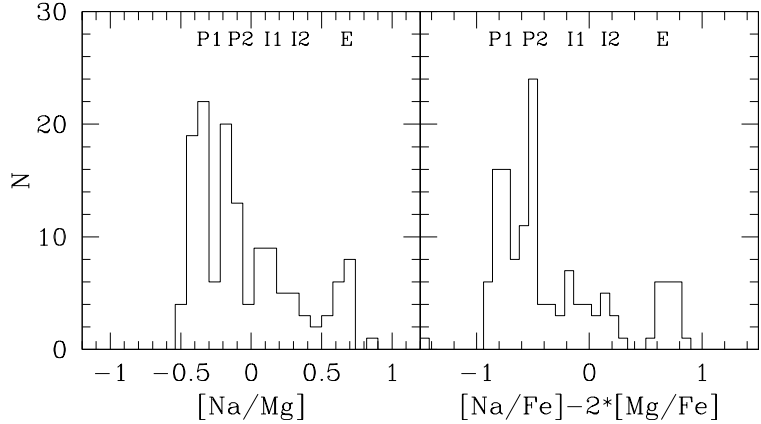}
\caption{Left panel: distribution function of the [Na/Mg] ratio for RGB stars 
in NGC~2808. In the right panel is plotted the distribution of the quantity
[Na/Fe]-2[Mg/Fe] for the same stars. The five peaks corresponding to the five
groups along the Na-Mg anticorrelation are labeled.}
\label{f:gruppiG2}
\end{figure*}

To check this appearence we examine the distribution function of the [Na/Mg]
ratio in the left panel of Fig.~\ref{f:gruppiG2}.

The distribution presents four clear peaks, with the group I2 seen as a tail
rightward of the I1 groups, because these two groups have quite
similar Na content, but Mg abundances decreasing by about 0.2 dex from I1 to I2
(see Fig.~\ref{f:gruppi}. A clearer representation of the five groups is given
by the right panel of Fig.~\ref{f:gruppiG2} where we plot the distribution
function relative to the quantity [Na/Fe]-2[Mg/Fe] to account for the fact that
the spread in Na is almost twice the spread in Mg. This is a sort of
pseudo-ratio, in analogy with pseudo-colours used in photometry 
\citep[see e.g.][and references therein]{mil15}.

In this case all five groups are clearly separated, and this plot is a mirror
image of the Fig.~\ref{f:histonao28}, assessing on a more robust statistical
basis the presence of five groups of stars with distinct abundances of
proton-capture elements on the RGB of NGC~2808.

We indicate with different colours in Fig~\ref{f:gruppi} these components: blue
(P1), green (P2), red (I1), orange (I2), and black (E)\footnote{Star 46580, with
[Na/Fe]=+0.212 dex and [Mg/Fe]=+0.268, may have a somewhat more uncertain
definition, as it is located in an intermediate position between the P2 and I1
groups. Looking at the ratios of the other elements, we assigned this star to
the P2 component. Star 55822, with no Mg determination, has [O/Fe]$=-0.644$ dex,
[Na/Fe]$=+0.608$ dex, and it is very likely that this object belongs to the E
component. Conservatively, we did not assign this star to any of the five group.
None of our results is however affected by these two objects.}.
Mean abundances for each group are listed in Tab.~\ref{t:mediet28}, where Al
abundances available for 31 giants with UVES spectra are from \citet{carr14},
with atmospheric parameters already on the present homogeneous scale.

\setcounter{table}{6}
\begin{table*}
\centering
\caption[]{Average abundances of proton-capture elements in the five groups on
the RGB of NGC~2808.}
\begin{tabular}{lccccc}

\hline
el.       &    P1             &    P2            & I1		   & I2 	       & E		   \\
\hline
          & n avg. rms        & n avg. rms       & n avg. rms	   & n avg. rms        & n avg. rms	   \\
$[$Fe/H$]$& 46 $-$1.128 0.028 &40 $-$1.134 0.026 &21 $-$1.130 0.024& 12 $-$1.137 0.028 & 20 $-$1.115 0.019 \\
$[$O/Fe$]$& 42   +0.308 0.058 &35   +0.154 0.090 &15 $-$0.216 0.124& ~7 $-$0.447 0.217 & 17 $-$0.656 0.161 \\
$[$Na/Fe$]$&46 $-$0.005 0.067 &40   +0.188 0.068 &21   +0.378 0.034& 12   +0.414 0.072 & 20   +0.592 0.112 \\
$[$Mg/Fe$]$&46   +0.384 0.041 &40   +0.346 0.035 &21   +0.274 0.044& 12   +0.127 0.028 & 20 $-$0.050 0.050 \\
$[$Al/Fe$]$&12   +0.065 0.062 &~9   +0.273 0.132 &~4   +1.006 0.118& ~2   +1.079 0.108 & ~4   +1.218 0.089 \\
$[$Si/Fe$]$&46   +0.265 0.026 &40   +0.262 0.026 &21   +0.309 0.026& 12   +0.346 0.038 & 20   +0.390 0.036 \\
$[$Na/Mg$]$&46 $-$0.388 0.059 &40 $-$0.158 0.051 &21   +0.104 0.057& 12   +0.287 0.069 & 20   +0.642 0.083 \\

\hline        
\hline
\end{tabular}
\label{t:mediet28}
\end{table*}

\begin{figure}
\centering 
\includegraphics[scale=0.42]{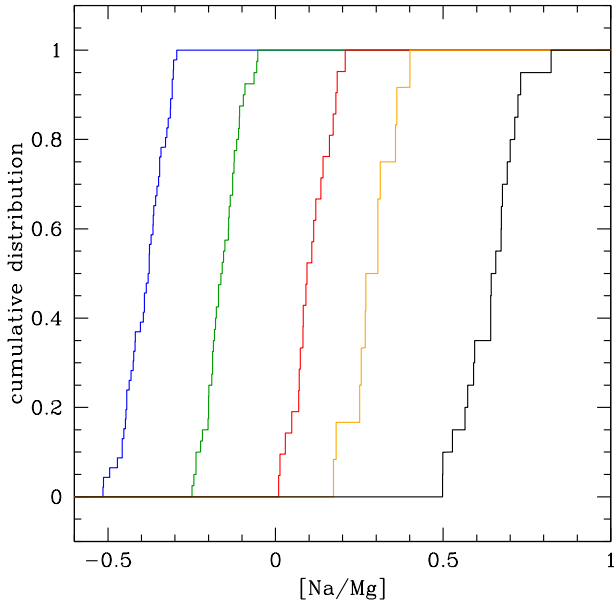}
\caption{Cumulative distribution of the [Na/Mg] ratio in the five groups of RGB
stars using the same colour-coding of Fig.~\ref{f:gruppi}: blue, green, red,
orange, and black indicate groups P1, P2, I1, I2, and E, respectively.}
\label{f:distrNAMG}
\end{figure}

The fraction of RGB stars in each group is P1$=33\pm 5\%$, P2$=29\pm 5\%$
I1$=15\pm 3\%$, I2$=9\pm 2\%$, and E$=14\pm 3\%$. 
All groups are well defined and distinct in the Na-Mg plane. In
Fig.~\ref{f:distrNAMG} we show how different they are by plotting the
cumulative  distributions of the [Na/Mg] ratio for the five components, using
the same colour-coding of Fig.~\ref{f:gruppi}.

To put on a quantitative foundation these differences, we test with a
Kolmogorov-Smirnov test the null hypothesis that any pair of the distributions
are extracted from the same parent distribution in [Na/Mg]. This hypothesis is
safely rejected in all cases, with the probabilities of rejecting it by mere
chance given in the last column of Tab.~\ref{t:cfrmean28}.

\begin{table*}
\centering
\caption[]{t-values, degrees of freedom and two-tailed probability values for
the  Student t-test on the mean abundance ratios for the five groups in NGC2808.
In the  last column the probabilities of a K-S test on cumulative distributions
of [Na/Mg] are also listed.}
\setlength{\tabcolsep}{1.4mm}
\begin{tabular}{lrrrrrrrr}

\hline
groups&[Fe/H]&      [O/Fe]       &     [Na/Fe]       &     [Mg/Fe]       &    [Al/Fe]        &      [Si/Fe]      &             [Na/Mg] &Prob(K-S)\\

\hline        
P1-P2 & 1.03 &  	     8.73& 		13.22&  	     4.64&		 4.38&  	     0.53&  		19.39  &$3.5\times 10^{-20}$ \\
d.f.  &   84 &  	       75& 		   84&  	       84&		   19&  	       84&  		   84  & 	 \\
p     & 0.306&  	     0.00& 		 0.00&$1.2\times 10^{-5}$&$3.2\times 10^{-4}$&  	    0.598&  		 0.00  & 	 \\
\hline        
P1-I1 &  0.30&  	    15.76& 		31.00&  	     9.69&		15.26&  	     6.43&  		32.41  &$6.0\times 10^{-14}$	 \\
d.f.  &    65&  	       55& 		   65&  	       65&		   14&  	       65&  		   65  & 	 \\
p     & 0.765&  	     0.00& 		 0.00&  	     0.00&		 0.00&$2.0\times 10^{-8}$&  		 0.00  & 	 \\
\hline        
P1-I2 &  0.99&  	     9.15& 		18.21&  	    25.46&		12.93&  	     6.97&  		31.06  &$1.5\times 10^{-9}$	 \\
d.f.  &    56&  	       47& 		   56&  	       56&		   12&  	       56&  		   56  & 	 \\
p     & 0.326&  	     0.00& 		 0.00&  	     0.00&$2.0\times 10^{-8}$&$1.5\times 10^{-5}$&  		 0.00  & 	 \\
\hline        
P1-E  &  2.19&  	    24.06& 		22.18&  	    34.15&		24.04&  	    14.02&  		50.25  &$1.6\times 10^{-13}$ 	 \\
d.f.  &    64&  	       57& 		   64&  	       64&		   14&  	       64&  		   64  & 	 \\
p     & 0.032&  	     0.00& 		 0.00&  	     0.00&		 0.00&  	     0.00&  		 0.00  & 	 \\
\hline        
P2-I1 &  0.60&  	    10.44& 		14.54&  	     6.50&		 9.96&  	     6.71&  		17.67  &$32.3\times 10^{-13}$ 	 \\
d.f.  &    59&  	       48& 		   59&  	       59&		   11&  	       59&  		   59  & 	 \\
p     & 0.551&  	     0.00& 		 0.00&$2.0\times 10^{-8}$&$7.7\times 10^{-7}$&  	     0.00&  		 0.00  & 	 \\
\hline        
P2-I2 &  0.33&               7.20&               9.66&              22.36&               9.14&               7.17&  		20.71  &$2.7\times 10^{-9}$ 	 \\
d.f.  &    50&	               40&	           50&                 50&                  9&                 50&  		   50  & 	 \\
p     & 0.743&$1.0\times 10^{-8}$&               0.00&               0.00&$7.5\times 10^{-6}$&               0.00&  		 0.00  & 	 \\
\hline        
P2-E  &  3.21&  	    19.33&              14.82&  	    31.74&              15.10&              14.16&  		39.53  &$5.6\times 10^{-13}$ 	 \\
d.f.  &    58&  	       50&                 58&  	       58&                 11&                 58&  		   58  & 	 \\
p     & 0.002&  	     0.00&               0.00&  	     0.00&$1.0\times 10^{-8}$&               0.00&  		 0.00  & 	 \\
\hline        
I1-I2 &  0.73&  	     2.62&               1.63&  	    11.71&               0.76&               3.00&               7.79  &$7.0\times 10^{-6}$ 	 \\
d.f.  &    31&  	       20&                 31&  	       31&                  4&                 31&                 31  & 	 \\
p     & 0.471&  	    0.016&              0.113&  	     0.00&              0.490&              0.005&$1.0\times 10^{-8}$  & 	 \\   
\hline        
I1-E  &  2.22&  	     8.71&               8.19&  	    21.98&               2.87&               8.22&              24.08  &$3.3\times 10^{-10}$ 	 \\
d.f.  &    39&  	       30&                 39&  	       39&                  6&                 39&                 39  & 	 \\
p     & 0.032&  	     0.00&               0.00&  	     0.00&              0.028&               0.00&               0.00  & 	 \\
\hline        
I2-E  &  2.41&  	     2.30&               7.96&  	    12.83&               1.57&               3.23&              13.04  &$1.0\times 10^{-7}$ 	 \\
d.f.  &    30&  	       22&                 30&  	       30&                  4&                 30&                 30  & 	 \\
p     & 0.022&  	    0.031&$1.0\times 10^{-8}$&  	     0.00&              0.192&              0.003&               0.00  & 	 \\

\hline        
\hline
\end{tabular}
\label{t:cfrmean28}
\end{table*}

\section{Discussion}

The homogeneous re-analysis of 140 red giants allows to uncover five different
populations, with distinct chemical composition, on the RGB of NGC~2808, thanks
to large statistics and a significant decrease of internal errors on abundances.
We can look now in more detail to the properties of these components.

\subsection{Chemical tagging of the five groups on the RGB}

\begin{figure}
\centering 
\includegraphics[scale=0.45]{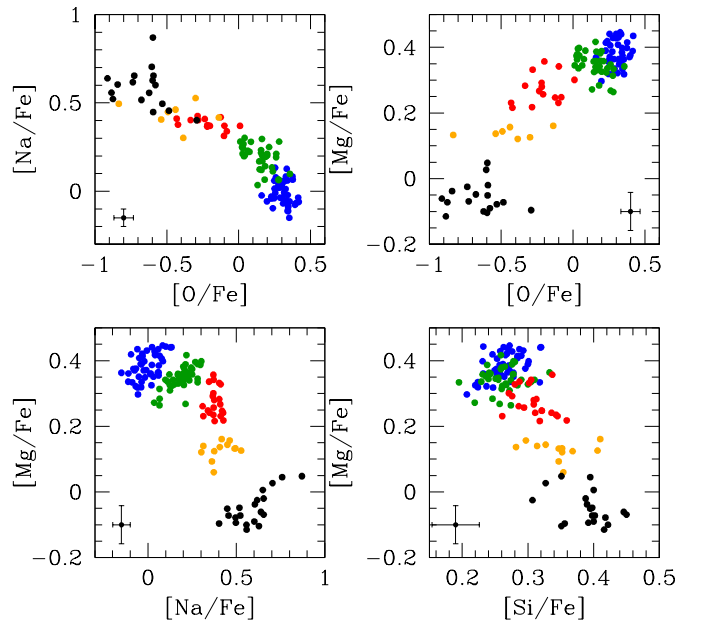}
\caption{From bottom-left and clockwise: Mg-Na, Na-O anticorrelations, Mg-O
correlation and Mg-Si anticorrelation among RGB stars in NGC~2808. Different
colours indicate the five populations as defined in the Na-Mg plane. In each
panel the star-to-star error bars are indicated.}
\label{f:lightt28}
\end{figure}

The separation of RGB stars into five groups has been assessed from both the 
classical Na-O and the Na-Mg anticorrelations. The last was used to operatively
define the five populations, and with this criterion we plotted with different
colour stars on the anticorrelations and the correlation among proton-capture
elements Na, O, Mg, Si in Fig.~\ref{f:lightt28}.

The classification with the [Na/Mg] results efficient in ranking the five
populations in all the other abundance planes, ordering the P1, P2, I1, I2, and 
E groups with decreasing Mg, and O abundances and increasing Na, Si abundances.
The boundaries of the populations are however more neatly defined in the Na-Mg
plane. Some ``blurring" is present at the border between the primordial
populations  P1 and P2 when O is involved, although the bulk of stars is well
separated. The smearing of stars in P1 and P2 appears to be more severe with the
Si abundances (bottom-right panel in Fig.~\ref{f:lightt28}). However, this is
expected because the production of $^{28}$Si is not a main process in the
proton-capture reactions, but only a leakage from the Mg-Al cycle \citep{kar03}.
When the bulk of abundances is still primordial, with the large
overabundances of $\alpha-$elements from type II supernovae nucleosynthesis
still in place, a slight enhancement in Si from $^{27}$Al($p$,$\gamma$)$^{28}$Si
is not clearly discernible. However, as more severe changes in composition from
the action of hot H-burning start to appear among stars, also the modification
to the Si level becomes more evident, as shown by stars of the I1, I2, and E
populations.

Abundances of Mg and Al for the 31 giants with UVES spectra were derived in 
\citet{carr14} using the present homogeneous scale. The Mg-Al anticorrelation
shows three clumps of stars with distinct chemical composition and the three
discrete populations account for $68\pm15\%$, $19\pm8\%$, and $13\pm4\%$ of
stars, for the P,I,E components, respectively.
A comparison with \citet{carr14} shows the P stars along the Mg-Al
anticorrelation are split among the present P1 and P2 groups, the I stars in the
I1 and I2, and finally the E stars in \citet{carr14} are found to belong here
all to the E population (see Tab.~\ref{t:mediet28}) with severe Mg depletions.
The fractions derived independently along the Mg-Al anticorrelation are in
excellent agreement (within the associated Poisson errors) with number counts
for the total P1+P2 (62\%), I1+I2 (24\%), and E (14\%) populations as defined
here from the Na-Mg anticorrelation.

A more quantitative description of the chemical composition in the five
populations od RGB stars is summarised in Tab.~\ref{t:cfrmean28}. We used
Student's and Welch's tests to determine if two sets of data are significantly
different from each other, for every one of the 10 combinations of the five
groups. We tested the null hypothesis that any pair of components are extracted
from a distribution having the same mean [element/Fe], for [Fe/H], [O/Fe],
[Na/Fe], [Mg/Fe], [Al/Fe], [Si/Fe], and [Na/Mg]. For any combination and element
we list the t-value, the number of degrees of freedom, and the two-tailed 
probability. 

The first consideration is that the metallicity [Fe/H] is the same for almost
all populations. Only the value of the E population seems to be statistically
different from the metallicity of the primordial and intermediate components.
This result is in qualitative agreement with the findings of \citet{bra10}
in a large set of GCs and in NGC~2808 in particular, and with the
expectation that the extreme population E must be enhanced in helium. By
consequence for a fixed global metallicity a decrease in the H abundance is
predicted, so that the ratio [Fe/H] is expected to slightly increase.

Mean O abundances differ among all groups and despite a couple of mutual 
interlopers (Fig.~\ref{f:lightt28}, top left panel) the average abundances of
populations P1 and P2 are statistically different.
For Na, average abundances do not significantly differ only among the two 
intermediate
populations I1 and I2. The reality of this feature is also confirmed by the fact
that their mean abundances of Al (that correlates with Na) are also not
significantly different. As already noted by visual inspection of
Fig.~\ref{f:lightt28}, the mean [Si/Fe] ratio is similar for the primordial
groups P1 and P2, but it differ in any other case. Finally, the [Mg/Fe] and
[Na/Mg] ratios are found to be statistically different in all populations at a
very high level of confidence.

Each of the five populations selected from the Na-Mg plane seems to be a
$single$ stellar population. Following the approach used by \citet{gra11}
for RHB stars, typical (anti)correlations would be observable if a single
group were composed by more than one population with homogeneous composition.
We verified this issue by looking at the presence of a Mg-Na anticorrelation or
a Mg-O correlation within each group. In each case we found either the reverse
relation (Mg correlated with Na or anti-correlated with O) or the correct
relation was found not to be statistically robust. Moreover, in each population
the [Na/Mg] ratio presents a single peaked distribution.

\begin{figure}
\centering 
\includegraphics[scale=0.40]{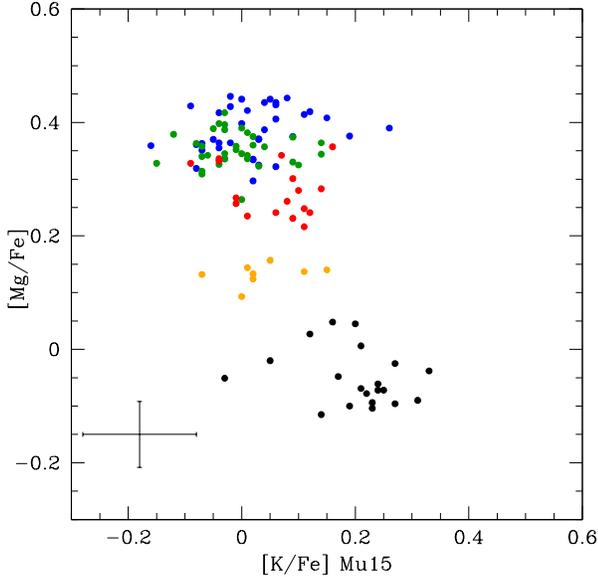}
\caption{The ratio [Mg/Fe] as a function of the [K/Fe] ratio from \citet{muc15}
in giants of NGC~2808, using the same colour-coding of 
Fig.~\ref{f:gruppi} for stars of populations P1, P2, I1, I2, and E.}
\label{f:kmg28}
\end{figure}

Recently \citet{muc15} (Mu15) derived potassium abundances for 116
giants of our sample in NGC~2808 (excluding three stars flagged as possible
binaries). They found that this GC is the second one after NGC~2419
\citep{muc12,coh12} where a Mg-K anticorrelation is clearly 
observed.
In Fig.~\ref{f:kmg28} we show our [Mg/Fe] ratios as a function of their [K/Fe] 
values, putting the Mg-K anticorrelation on a larger statistical basis and
confirming that K in a few GCs participates to the same self-enrichment network
due to proton-capture reactions.
A more precise comparison with the results of \citet{muc15} is
hampered by the fact that they used for NGC~2808 the older temperature scale by
\citet{car06,car09a}. However, there is a way to bypass this obstacle.

The Si-Al correlation and the mirroring Mg-Si anticorrelation, observed
in NGC~2808 and other GCs \citep{yon05,car09a}, are 
thermometers indicating that the temperature of the H-burning generating this
pattern exceeded T$_6 \sim 65$ K \citep{arn99}, because at this
temperature the reaction $^{27}$Al($p$,$\gamma$)$^{28}$Si becomes dominant over 
$^{27}$Al($p$,$\alpha$)$^{24}$Mg. However, this is only a lower limit to the
temperature, meaning that from now on the leakage on $^{28}$Si from the Mg-Al
cycle is possible. A more elevated temperature regime is required to explain the
changes in the K content as observed in NGC~2419 \citep{ven12}. In their
attempt to account for the extreme K-enhanced, Mg-depleted stars observed by
\citet{muc12}, Ventura et al. postulated that at temperatures above
10$^8$ k is
activated a process of proton-capture on $^{36}$Ar that produces potassium and 
shifts the equilibrium among various species toward heavier
nuclei like K, Ca, and Sc, in particular when low metallicity models are
considered. The effect is relevant on K (see Fig.~\ref{f:kmg28})
and on Sc, because they are much less abundant that Ca: a fraction of the
original Ar is sufficient to provide large abundance variation. 
Different is the case of Ca, since the original content of this element is 
larger than that of K and much more abundant than Sc.

\begin{figure}
\centering 
\includegraphics[bb=19 146 472 708, clip, scale=0.52]{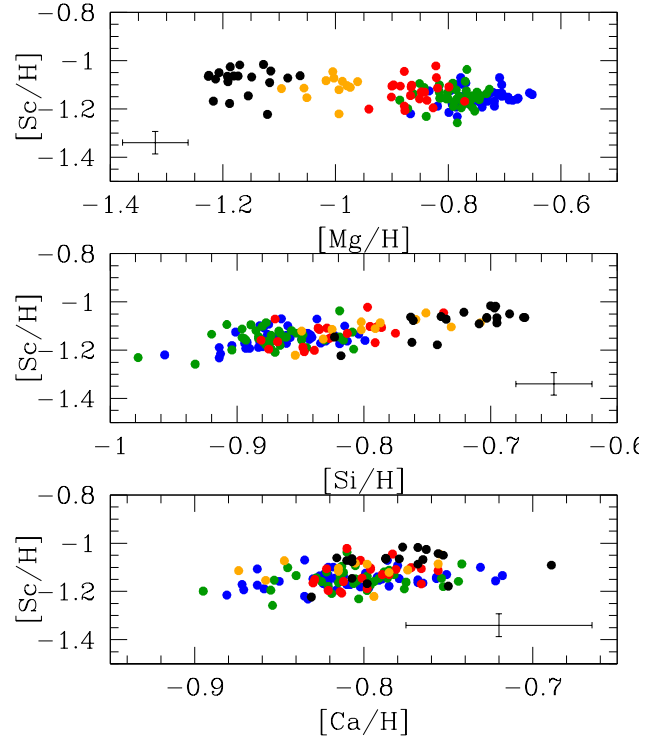}
\caption{From top to bottom, Sc as a function of the abundances of Mg, Si,
and Ca in our sample. Internal error bars are shown in each panel.}
\label{f:sct28h}
\end{figure}

Hence, if this process were at work also in (at least part of) first generation 
stars at the higher metallicity of NGC~2808, we could test it by using our
homogeneous set of abundances without being afraid of possible offests between
different analyses, as for K, since with the same scale we derived in the
present work abundances of Sc, Mg, Si, and Ca.
The relations between these elements are shown in Fig.~\ref{f:sct28h}. The
probability that Sc and Mg are not anticorrelated is P=$1.0 \times 10^{-8}$
whereas that of Sc and Si lacking a correlation is P=0.0. For the correlation
Sc-Ca we found P=$3.5 \times 10^{-5}$, with 140 d.o.f.
Finally, the correlations
of Sc and Si with K are both found to be statistically significant.
We may conclude that clear evidence does exist for the action of 
proton-capture occurring at a particularly high temperature regime in a part of
stars that contributed to the pollution of intracluster gas at the epoch of the
formation of the following generation(s) of stars. A clue on the origin of
this processed material is however hard to assess since such temperatures are
encountered in the two main candidate polluters proposed for GCs, namely massive
AGBs and the most massive main-sequence stars \citep[see][]{pra07}.

\begin{figure}
\centering 
\includegraphics[scale=0.40]{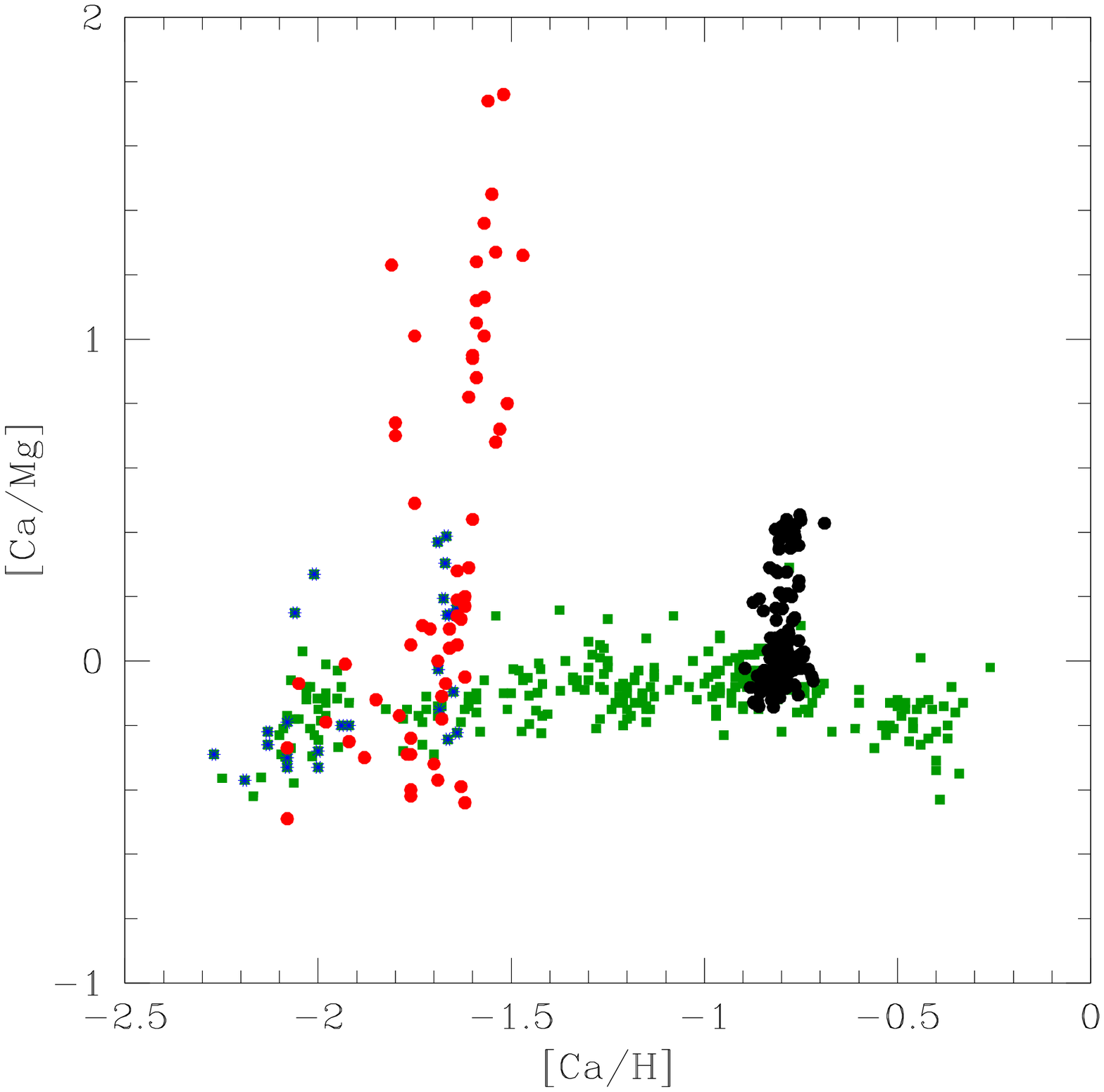}
\caption{The ratio [Ca/Mg] as a function of the [Ca/H] ratio for giants in
NGC~2419 \citep[red filled circles]{muc12,coh12},
in about 280 RGB stars in more than 20 
GCs \citep[green squares]{car09a,car10b,car11,car13a,car14,car15,car10c}, and
giants in NGC~2808 (present work; black circles). Blue asterisks indicate M~15
and NGC~4833.}
\label{f:versione28}
\end{figure}

The cases of NGC~2419 and NGC~2808 point out, albeit at a different degree, the 
key r\^ole of the Mg-depletion as useful indicator of a particular environment where the
network of proton-capture reactions operate at unusually high temperature.
We may then confirm the plot in Fig.~\ref{f:versione28}, first introduced in
\citet{car13b}, as a powerful diagnostic to uncover GCs where this
extreme processing occurred. In this plot we compare one of the element most
affected by this advanced nucleosynthesis, Mg, with one of the least varied, Ca.
Using as a reference the homogeneous analysis of more than 280 RGB stars with
UVES spectra in 23 GCs (green squares) it is possible to pick up clusters were
signatures of such a particular nucleosynthesis are visible. Apart from the
exceptional case of NGC~2419, only NGC~2808, NGC~4833 and M~15 stand out, with
the last two GCs being much more metal-poor. 

\subsection{Comparison with photometry}

\citet{mil15} identified at least five clumps of stars on
the RGB in NGC~2808 from precise UV photometry with $HST$ and colour or
pseudo-colour indexes purposedly tailored to maximize the separation among
photometric sequences due to the absorption of molecular bands like OH, NH, CN,
and  CH of elements involved in proton-capture reactions. The counterparts of
these discrete populations are also found among MS stars, although less clearly
since molecular bands become less prominent in warmer stars. Using 32 giants in
their  set with abundances from \citet{car04,car06} and \citet{carr14}
Milone et al. showed that their populations from B to E present signatures  of
increasingly processed composition, namely depletions in O and Mg and
enhancements in Na and Al. The metallicity and composition of their population A
is unknown, lacking stars with spectroscopy, but they inferred a primordial
abundance pattern in this group.

\begin{figure}
\centering 
\includegraphics[scale=0.40]{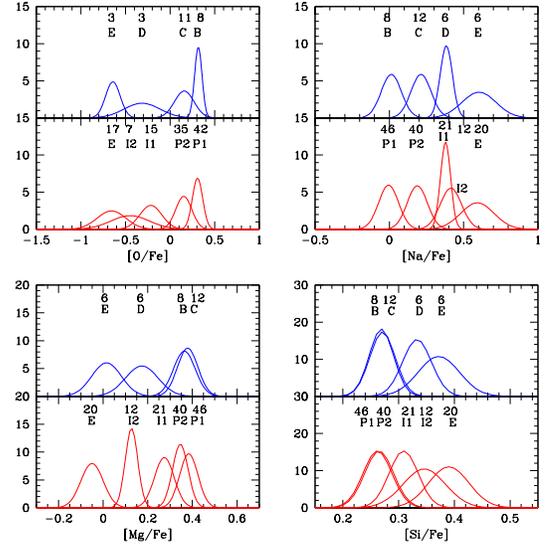}
\caption{Comparison of mean abundance ratios for the photometric populations 
B,C,D,E identified by \citet[upper half of each panel]{mil15}) with
average ratios of the five populations P1,P2,I1,I2,E (lower half of each 
panel). Each population is represented by a gaussian centered at the mean value
and with $\sigma$ given by the $rms$ scatter of the mean ratio.}
\label{f:cfrpop28}
\end{figure}

A direct comparison is difficult because the photometric catalogue are not yet
published and usually $HST$ fields are taken on the cluster centre, with
scarce overlap with the pointings of multiobject spectrographs like FLAMES.
However, we can use the 32 stars in common to bridge the gap and compare the
five groups detected photometrically to the five populations with distinct
composition uncovered in the present work.
For each group of stars associated by \citet{mil15} to populations
B,C,D,E we recomputed the average and $rms$ scatter of the [O/Fe], [Na/Fe],
[Mg/Fe], and [Si/Fe] ratios using the presently derived new abundances. Al was 
not considered, since only a few stars from \citet{carr14}
are in the photometric sample (all in the primordial regime,
three in group B and two in C). We used these values simply to locate the
position of the photometric groups in the upper half of the four panels in
Fig.~\ref{f:cfrpop28}, representing each group as a gaussian centered at the
elemental mean ratio, and having the corresponding $rms$ scatter as $\sigma$. 

In the bottom half of each panel we plotted the position of our five populations
using the values in Tab.~\ref{t:mediet28}. This comparison shows that the match
is satisfactory but not perfect. The cross-identification of a few populations,
namely the extreme components, seems to be firmly established.
The two groups P1 and P2 with primordial or almost primordial composition
correspond to the populations B and C, respectively, for each of the four
elements. We may conclude that the parent populations of the two groups of RHB
stars in NGC~2808 are then traced on the RGB using both photometry and
spectroscopy. Also the extreme population is easily (and unambigously) 
identified in the present work and in the $HST$ photometry by Milone and
collaborators.  

Some problems seem to arise when dealing with the intermediate components. In
the panels of Fig.~\ref{f:cfrpop28} for O, Mg, and Si the photometric component
D looks more or less coincident with our population I2: in these plots seems
difficult to trace among the photometric groups the counterpart of I1, the other
population with intermediate composition. On the other hand, when looking at Na,
the photometric group D seems to be associated more to the I1 population than to
I2. However, this mismatch may actually be only apparent, because in Fig. 3 of
\citet{mil15} group D shows  a large spread in the index $\Delta_{\rm
F336W,F438W}$, with a tail visible at low values. Since this index is sensitive
to the abundances of light elements (C, N), it is likely that also the group D
is not a simple stellar population, as already stated for groups B and C by
Milone et al.

\begin{table*}
\centering
\caption[]{Stellar populations in NGC~2808}
\begin{tabular}{lcccl}

\multicolumn{5}{c}{spectroscopy}      \\

\hline

Na-O RGB  &           P           &            I         &      E      &           \\
fraction  &       $45\pm6\%$      &   $36\pm6\%$         & $18\pm4\%$ & this work \\
          &                       &                      &            &           \\
Na-Mg RGB &  P1~~       ~~~~~P2    &     I1~~    ~~~~~~~I2    &     E      &           \\
fraction  & $33\pm5\%$ ~~$29\pm5\%$ & $15\pm3\%$ ~~$9\pm2\%$ & $14\pm3\%$ & this work \\  
          &                       &                      &            &           \\
Mg-Al RGB &           P           &            I         &      E      &           \\
fraction  &       $68\pm15\%$     &   $19\pm8\%$         & $13\pm4\%$ & Carretta 2014 \\
          &                       &                      &            &           \\
from O RGB&         O-normal      &     O-poor           &   super O-poor      &           \\
fraction  &       $61\pm17\%$     &   $22\pm4\%$         & $17\pm4\%$ & Carretta+ 2006 \\
          &                       &                      &            &           \\
Na-O HB   &         RHB           &     BHB              &            &           \\
          &    O-rich/Na-poor     &  O-poor/Na-rich      &            & Gratton+ 2011 \\
          &    moderate Na-O      &                      &            &           \\
          &    anticorrelation    &                      &            &           \\
          &                       &                      &            &           \\
Na-O HB   &         RHB           &     BHB              &            &           \\
          &    lower Na           &  higher Na           &            & Marino+ 2014 \\
          &    bimodal Na         &  higher He           &            &           \\
          &                       &                      &            &           \\
K in RGB  &         Primordial    &     Intermediate     &   Extreme  &           \\
          &       low K           &   intermediate K     & high K     & Mucciarelli+ 2015 \\
          &                       &                      &            &           \\

\hline
\hline

\\
\multicolumn{5}{c}{photometry}      \\

\hline
MS        &         rMS              &         mMS             &     bMS       &           \\
fraction  &   $62\pm2\%$             &      $24\pm2\%$         &    $14\pm3\%$ &  Milone+ 2012   \\
          &                          &                         &               &           \\
AGB       &     AGB(I)               &        AGB(II)          &    AGB(III)   &           \\
fraction  &   $49\pm11\%$            &      $22\pm6\%$         &    $29\pm8\%$ &  Milone+ 2015   \\
          &                          &                         &               &           \\
AGB       &     A  B  C              &         D               &    E          &           \\
fraction  &   A $5.8\pm0.5\%$        &      $31.3\pm1.3\%$     &$19.1\pm1.0\%$ &  Milone+ 2015   \\
          &   B $17.4\pm0.9\%$       &      	               &               &                 \\
          &   C $26.4\pm1.2\%$       &      	               &               &                 \\
          &                          &                         &               &           \\
estimate  &    A  B C                &         D               &     E         &           \\
from MS   & Y=0.246 0.278 0.280      &       Y=0.329           &    Y=0.384    &  Milone+2015 \\
from RGB  & Y=0.243 0.278 0.280      &       Y=0.318           &    Y=0.367    &              \\
          &                          &                         &               &           \\
from HB   &    RHB                   &         BHB             &     EHB       &           \\
fraction  &   $41\pm3\%$             &     $39\pm3\%$          & $9\pm1\%$     &  Dalessandro+ 2011 \\
estimate  & Y=0.248                  &       Y=0.30            &    Y=0.40     &                    \\
          &                          &                         &               &           \\
from HB   &    RHB                   &         EBT1            &     EBT2      &           \\
fraction  &   $41\pm3\%$             &     $34\pm2\%$          & $14\pm1\%$    &  Iannicola+ 2009 \\
          &                          &                         &               &           \\
from HB   &    RHB                   &         EBT1            &     EBT2      &           \\
fraction  &   $50\%$                 &     $30\%$              & $20\%$        &  D'Antona+ 2005 \\
estimate  & Y=0.24                   &       Y=0.26-0.29       &    Y=0.40     &                    \\
          &                          &                         &               &           \\
from HB   &    RHB                   &         EBT1            &     EBT2      &           \\
fraction  &   $46\pm10\%$            &     $35\pm10\%$         & $10\pm5\%$    &  Bedin+ 2000 \\

\hline
\hline
\end{tabular}
\begin{list}{}{}
\item Notes: residual components on the HB are as follows:
\item EBT3 $9\pm5\%$ (Bedin+2000)
\item BHk $9\pm1\%$ (blue hook stars, Dalessandro+2011)
\item EBT3 $11\pm1\%$ (Iannicola+2009)
\end{list}
\label{t:tabfig}
\end{table*}

In Tab.~\ref{t:tabfig} we summarized recent results from spectroscopy and
photometry to estimate the number of components detected in different
evolutionary phases in NGC~2808, and the fractions of stars attributed to each
one. Despite the variety of methods, and evolutionary phases sampled a few facts
emerge. 

\begin{figure}
\centering 
\includegraphics[scale=0.42]{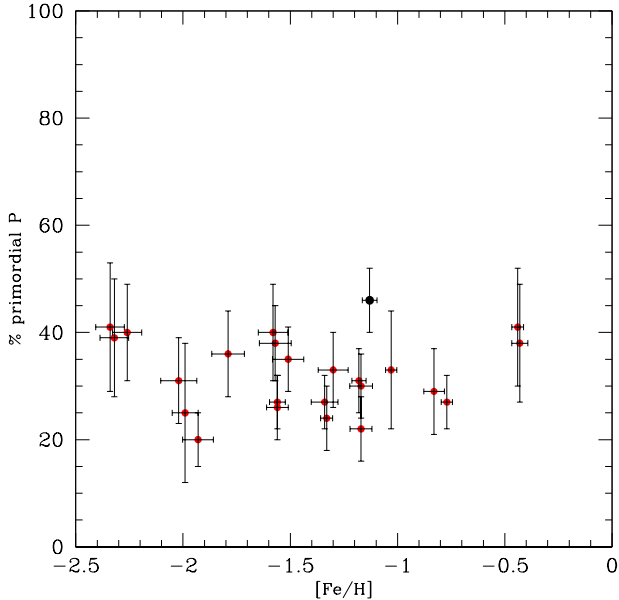}
\caption{Fractions of first generation stars in 24 GCs of our FLAMES survey as a
function of the metallicity [Fe/H]. Attached error bars in fractions are from 
Poisson's statistics. For metallicity we used systematic errors (from
Tab.~\ref{t:sensitivityt28} and similar tables in the other papers). NGC~2808 is
indicated with a larger black symbol.} 
\label{f:piefe}
\end{figure}

First, NGC~2808 hosts the largest fraction of stars with primordial composition
observed among massive Milky Way GCs up to date: about 50\% of the currently
observed stellar population belongs to the original generation formed in this
object. This feature is clearly appreciated in Fig.~\ref{f:piefe}, based on the
now fully homogeneous analysis of Na, O abundances for 1598 red giants in 24 GCs
from our FLAMES survey, where we plot the fraction of stars in the P component
as a function of metallicity.
The estimates for NGC~2808 range from about 41\% from number count on the RHB
\citep{ian09,dal11} up to the maximum value of 68\% 
derived from the Mg-Al anticorrelation in RGB stars \citep{carr14}. Even such a
high value seems however confirmed by number counts on the red MS 
\citep{mil12b}. Since on average the stellar primordial content in GCs is much lower 
(about 30\% as estimated from the Na-O anticorrelation) an open question is how
and why NGC~2808 retained such a higher fraction.

The second feature inferred from Tab.~\ref{t:tabfig} is that a gross division
into three components seems to be accounted for by most analysis. For the
intermediate group, with clearly modified but not extreme chemical composition,
a fraction ranging from one fifth up to one third is observed by most studies.
The extreme component is also traced (whenever accessible to observations) by 
most works, with reassuringly similar estimates of the fraction of stars.

However, a third fact is assessed by recent studies, although hints were already
suggested by early investigations of the multimodal HB in NGC~2808: a finer
subdivision of components is corroborated from high precision photometry and
spectroscopy. The cross match between abundances and colour indexes is usually
good and mutually suggests that what we are observing are real stellar
populations forming discrete sequences or groups.

\section{Summary and conclusions}

From high/intermediate spectra we reanalysed 140 individual RGB stars in
NGC~2808 using the same homogeneous procedures adopted for more than 2500 giants
in other 23 GCs.
We derived abundances of Fe, O, Na, Mg, Si, Ca, Sc, Ti, Cr, Mn, Ni.
Pure $\alpha-$elements and elements of the Fe-group are constant among multiple
populations in NGC~2808. On our high resolution UVES metallicity scale
\citep{car09c} we found a metal abundance for
NGC~2808 of [Fe/H]$=-1.129\pm0.005\pm0.034$ ($\pm$statistical
$\pm$systematic error) with $\sigma=0.030$ (31 stars).

We confirm that the Na-O anticorrelation in this cluster is one of the most
extended observed in GCs with no intrinsic spread in iron, in agreement with the
high cluster present day total mass and with the hot temperatures reached on the
blue HB.
According to the definition of \citet{car09b} NGC~2808 host the highest
fraction of the primordial, first generation stars detected spectroscopically.
The improved statistics and the smaller internal errors due to uncertainties in
the analysis uncover that stars along the Na-O anticorrelation seems clustered
in five groups.

The larger sample of stars with Na, Mg abundances allow to confirm that five
populations of RGB stars with distinct composition of proton-capture elements 
are located along the Na-Mg anticorrelation. Statistical tests provide robust
evidence of the different chemistry of these components that we called P1, P2,
I1, I2, and E in order of decreasing Mg and increasing Na abundances.
Large Mg depletions are observed in NGC~2808, with a fraction of about 14\% of
stars (the E component) having solar or even subsolar [Mg/Fe] ratios.

We confirm and extend in statistics the anticorrelation between Mg and potassium
abundances found by \citet{muc15} in this GC. Mg anticorrelated with
K and Si, as well as with Sc, is proof that proton-capture processes operated in
a very elevated temperature regime for H-burning in part of stars of the first 
generation in NGC~2808. 

The match of the five populations detected from spectroscopy and the five groups
recently discovered by \citet{mil15} with UV photometry is good but not
perfect. The two populations with almost primordial abundances are well
cross-identified, nicely tracing the progenitors of the two groups of RHB stars
individuated by \citet{gra11} and \citet{mar14}. There is also a 
definite agreement on the extreme component. 
The partial mismatch of the populations with intermediate composition confirm
the evidence that other discrete components may exist in NGC~2808, as suggested
by the photometry.

Apparently, current modelling seems to lag somewhat behind observations. The
strong depletions detected in NGC~2808 \citep[and the present analysis]{carr14}
offer a severe challenge to the main scenario of self-enrichment of GCs in early
epochs. Models of asymptotic giant branch (AGB) or SAGB stars with the proper
metallicity (Z=0.001) produce the required depletions in Mg \citep{ven13},
but not enough to match the reduction factors observed in NGC~2808. On
the other hand, the wind from model 60rD in the set of fast
rotating massive stars (FRMS) by \citet{dec07} seems to be able to
account for a large range of Mg depletions and Al enhancement, but at the price
of a strong destruction of Na, which is not observed in stars of NGC~2808 with
extreme composition \citep[this work and][]{carr14}.
For the time being, we note that \citet{bas15} 
showed how no one of the most fashionable models proposed for
polluting first generation stars is able to fit at the same time the abundance
pattern of normal GCs $and$ the extreme modifications observed in NGC~2808. Once
again, this cluster is a pivotal object to interpret the evidence of multiple
populations in GCs.

The discreteness of the different populations detected in NGC~2808 (see
Tab.~\ref{t:tabfig}) presents another level of challenge. In most scenarios 
\citep[see]{gra12} the observed abundance pattern only may be approximated by
diluting ejecta of favourite candidate polluters with not processed, primordial 
matter. It seems difficult to explain populations with discrete composition 
by imaging a formation history where bursts of star formation are followed by
multiple quiescent periods, whereas at the same time the abundances continue to
change from a population to the next, following an empirical dilution model.
To the difficulty of accounting for the origin and collection of pristine  gas
\citep[e.g.][]{der11}, one is forced to add mechanisms to selectively stop
the mixing and/or star formation several times in a few 10$^7$-10$^8$ yr at the
cluster formation.

A more definitive answer to the questions aroused by these recent discoveries in
NGC~2808 must necessarily await when photometric catalogues will be published
and the analysis of spectroscopic observations of stars in the photometric
groups completed by Milone, Marino and coworkers. 
Our group just gathered high resolution, high S/N spectra for about 100 giants in 
the present sample to measure Al abundances, already proved in NGC~6752
\citep{car12} and NGC~2808 \citep{carr14} to be very useful to
chemical tag discrete stellar populations on the RGB. With the full set of
elements involved in proton-capture reactions we will be able to perform a 
statistically 
sound cluster analysis and hopefully disclose other windows on this very
particular globular cluster.
                    
\acknowledgements
We thank Angela Bragaglia for valuable help and discussions
and Alessio Mucciarelli for providing his results for Potassium. This
publication makes use of data products from the Two Micron All Sky Survey, which
is a joint project of the University of Massachusetts and the Infrared
Processing and Analysis Center/California Institute of Technology, funded by the
National Aeronautics and Space Administration and the National Science
Foundation.  This research has been funded by PRIN MIUR 2010-2011, project ``The
Chemical and Dynamical Evolution of the Milky Way and Local Group Galaxies'' (PI
F. Matteucci) . This research has made use of the SIMBAD database, operated at
CDS, Strasbourg, France and of NASA's Astrophysical Data System.

\clearpage

\setcounter{table}{0}
\begin{table*}
\centering
\caption[]{Adopted atmospheric parameters and derived iron abundances in
NGC~2808. The prefix U indicates stars observed with UVES. The
complete table is available electronically only at CDS.}
\begin{tabular}{rccccrcccrccc}
\hline
Star   &  $T_{\rm eff}$ & $\log$ $g$ & [A/H]  &$v_t$	     & n  & [Fe/H]{\sc i} & $rms$ & n  & [Fe/H{\sc ii} & $rms$ \\
       &     (K)	&  (dex)     & (dex)  &(km s$^{-1}$) &    & (dex)	  &	  &    & (dex)         &       \\
\hline
7183	&4531 & 1.52 & $-$1.08 & 1.70 &  46 & $-$1.080 & 0.071 &  4  & $-$1.121 &0.257  \\  
7315	&4459 & 1.45 & $-$1.14 & 1.29 &  54 & $-$1.138 & 0.086 &  3  & $-$1.075 &0.078  \\  
U7536	&4306 & 1.18 & $-$1.11 & 1.55 &  89 & $-$1.109 & 0.079 & 10  & $-$1.125 &0.055  \\  
7558	&4711 & 1.87 & $-$1.13 & 1.64 &  47 & $-$1.129 & 0.086 &  4  & $-$1.106 &0.270  \\  
7788	&4510 & 1.53 & $-$1.15 & 1.44 &  20 & $-$1.144 & 0.058 &     &          &	\\  

\hline
\end{tabular}
\begin{list}{}{}
\item Note - This table is available in its entirety in a machine-readable form
in the online journal. A few lines are shown here for guidance regarding its
form and content.
\end{list}
\label{t:atmpart28}
\end{table*}

\setcounter{table}{3}
\begin{table*}
\centering
\caption[]{Abundances of proton-capture elements in stars of NGC~2808.
Upper limits (limO=0)and detections (=1) for O are flagged. The
complete table is available electronically only at CDS.}  
\begin{tabular}{rrccrccrccrcccc}
\hline
  star   &n   & [O/Fe] &  rms   & n & [Na/Fe] & rms   & n &[Mg/Fe]& rms    & n &[Si/Fe] & rms  &limO  \\ %
\hline       
7183 	 &  2 &  +0.232 & 0.033 & 4 &  +0.135 & 0.023 & 3 & +0.323 & 0.077 & 10 & +0.261 & 0.084 &  1	\\  
7315 	 &  2 &  +0.352 & 0.047 & 4 &$-$0.150 & 0.040 & 3 & +0.363 & 0.101 & 12 & +0.278 & 0.085 &  1	\\  
U7536	 &  2 &  +0.248 & 0.040 & 4 &  +0.060 & 0.017 & 3 & +0.355 & 0.095 &  8 & +0.260 & 0.039 &  1	\\  
7558 	 &  1 &  +0.392 &       & 3 &$-$0.076 & 0.067 & 2 & +0.370 & 0.009 & 11 & +0.249 & 0.072 &  1	\\  
7788 	 &  0 & 	&       & 2 &  +0.224 & 0.026 & 1 & +0.352 &	   &  8 & +0.236 & 0.066 &  1	\\  
\hline
\end{tabular}
\begin{list}{}{}
\item Note - This table is available in its entirety in a machine-readable form
in the online journal. A few lines are shown here for guidance regarding its
form and content.
\end{list}
\label{t:protont28}
\end{table*}

\setcounter{table}{4}
\begin{table*}
\centering
\caption[]{Abundances of $\alpha-$ and Fe-peak elements in stars of NGC~2808. 
The complete table is available electronically only at CDS.}
\begin{tabular}{rrccrccrccrccrcc}
\hline
   star & n &[Ca/Fe]& rms  & n & [Ti/Fe]~{\sc i} & rms & n &[Sc/Fe]~{\sc ii} &  rms &n &[Cr/Fe]~{\sc i} & rms &n &[Ni/Fe] & rms \\
\hline   
7183 	& 6 & +0.326 & 0.059 & 5 &  +0.228 & 0.071 &  6 &$-$0.066 & 0.071 & 5 &$-$0.052 & 0.078 &11 &$-$0.077 & 0.087 \\ 
7315 	& 7 & +0.317 & 0.113 & 5 &  +0.159 & 0.078 &  6 &  +0.039 & 0.151 & 3 &$-$0.053 & 0.138 & 7 &$-$0.123 & 0.031 \\ 
U7536	&13 & +0.318 & 0.069 &10 &  +0.189 & 0.047 &  8 &$-$0.045 & 0.103 &18 &$-$0.051 & 0.089 &30 &$-$0.066 & 0.064 \\ 
7558 	& 7 & +0.332 & 0.102 & 6 &  +0.215 & 0.072 &  6 &$-$0.057 & 0.072 & 5 &$-$0.061 & 0.081 &10 &$-$0.061 & 0.219 \\ 
7788 	& 1 & +0.339 &       & 2 &  +0.236 & 0.042 &  5 &  +0.050 & 0.062 & 5 &$-$0.024 & 0.121 & 2 &$-$0.064 & 0.007 \\ 
\hline
\end{tabular}
\begin{list}{}{}
\item Note - This table is available in its entirety in a machine-readable form
in the online journal. A few lines are shown here for guidance regarding its
form and content.
\end{list}
\label{t:alphat28}
\end{table*}

\setcounter{table}{5}
\begin{table*}
\centering
\caption[]{Abundances of Ti~{\sc ii}, Cr~{\sc ii}, and Mn in stars of NGC~2808 observed with UVES.}
\begin{tabular}{rrccrccrcc}
\hline
   star &  n & [Ti/Fe]~{\sc ii} & rms & n &[Cr/Fe]~{\sc ii} &  rms &n &[Mn/Fe] & rms \\
\hline   
 8739 	&  9 &   +0.208  & 0.094 &  14  &  $-$0.011 &  0.118 & 3 & $-$0.395 & 0.026 \\ 
38660 	&  8 &   +0.186  & 0.110 &  17  &  $-$0.048 &  0.174 & 3 & $-$0.355 & 0.009 \\ 
 8603	&  9 &   +0.195  & 0.091 &  14  &  $-$0.071 &  0.120 & 3 & $-$0.378 & 0.080 \\ 
10571	&  8 &   +0.197  & 0.072 &  12  &  $-$0.020 &  0.067 & 3 & $-$0.399 & 0.048 \\ 
30763	&  9 &   +0.200  & 0.070 &  14  &  $-$0.034 &  0.133 & 3 & $-$0.359 & 0.066 \\ 
\hline
\end{tabular}
\begin{list}{}{}
\item Note - This table is available in its entirety in a machine-readable form
in the online journal. A few lines are shown here for guidance regarding its
form and content.
\end{list}
\label{t:ticrmnt28}
\end{table*}

\end{document}